# Low Temperature Surface Formation of $NH_3$ and HNCO: hydrogenation of nitrogen atoms in CO-rich interstellar ice analogues

G. Fedoseev[1, a)], S. Ioppolo[2, 3], D. Zhao[1], T. Lamberts[1, 3] and H. Linnartz[1]

[1]*Sackler Laboratory for Astrophysics, Leiden Observatory, University of Leiden, PO Box 9513, NL 2300 RA Leiden, The Netherlands*

[2]*Division of Geological and Planetary Sciences, California Institute of Technology, 1200 E. California Blvd., Pasadena, California 91125, USA*

[3]*Institute for Molecules and Materials, Radboud University Nijmegen, PO Box 9010, NL 6500 GL Nijmegen, The Netherlands*

**ABSTRACT**

Solid state astrochemical reaction pathways have the potential to link the formation of small nitrogen-bearing species, like $NH_3$ and HNCO, and prebiotic molecules, specifically amino acids. To date, the chemical origin of such small nitrogen containing species is still not well understood, despite the fact that ammonia is an abundant constituent of interstellar ices toward young stellar objects and quiescent molecular clouds. This is mainly because of the lack of dedicated laboratory studies. The aim of the present work is to experimentally investigate the formation routes of $NH_3$ and HNCO through non-energetic surface reactions in interstellar ice analogues under fully controlled laboratory conditions and at astrochemically relevant temperatures. This study focuses on the formation of $NH_3$ and HNCO in CO-rich (non-polar) interstellar ices that simulate the CO freeze-out stage in dark interstellar cloud regions, well before thermal and energetic processing start to become relevant. We demonstrate and discuss the surface formation of solid HNCO through the interaction of CO molecules with NH radicals - one of the intermediates in the formation of solid $NH_3$ upon sequential hydrogenation of N atoms. The importance of HNCO for astrobiology is discussed.

## 1 INTRODUCTION

The detection of glycine, the simplest amino acid, in cometary samples recently returned to Earth by the STARDUST mission has boosted detailed investigations of the origin and fate of (pre)biotic molecules in the interstellar medium (ISM) (Elsila et al. 2009, Garrod 2013). However, although an increasing number of laboratory and theoretical studies show that complex species form in the solid phase, on the surface of icy grains, we still lack understanding of the complete surface formation pathways at play. The nitrogen chemistry of the ISM is particularly important within this context, because of its potential to reveal the formation routes of the simplest amino acids or their possible precursors. From the ~180 species

---

a) Author to whom correspondence should be addressed. Electronic mail: fedoseev@strw.leidenuniv.nl

unambiguously identified in the ISM, about one third contains nitrogen atoms, but only $NH_3$, XCN, and possibly $NH_4^+$ are identified as constituents of interstellar ices. Solid $NH_3$ is generally found with a typical abundance of 5% with respect to water ice toward low- and high-mass young stellar objects (YSOs) (Gibb et al. 2004, Bottinelli et al. 2010, Öberg et al. 2011). Solid isocyanic acid (HNCO) has not been identified in the solid phase yet, but its direct derivative, the cyanate ion ($OCN^-$), has been found in interstellar ices with abundances between 0.3-0.6% with respect to water ice. The assignment of solid $OCN^-$ is often attributed to either the entire, so called, XCN band or to a single component (2165 $cm^{-1}$) of the full band (van Broekhuizen et al. 2005). More recently, Öberg et al. (2011) found a correlation between CO and the XCN band that supports the identification of the latter as $OCN^-$. Another possible N-bearing component of interstellar ices is $NH_4^+$. Although the unambiguous assignment of $NH_4^+$ is still under debate (Gálvez et al. 2010), it can potentially be one of the carriers of the 5-8 μm bands, and its presence in interstellar ices is consistent with previously obtained laboratory results (Boogert et al. 2008). The existence of interstellar solid $NH_4^+$ is indeed constrained by the hypothesis that $NH_4^+$ helps to maintain charge balance between positive and negative ions within interstellar ices. Öberg et al. (2011) assigned $NH_4^+$ abundances of 2.3 and 4.3% with respect to water ice toward low- and high-mass protostars, respectively. The formation of $OCN^-$ and $NH_4^+$ is commonly associated with a later stage of molecular cloud evolution, when thermal processing of the ice by a newly formed protostar becomes important. $NH_3$ and HNCO are commonly considered the precursors of $NH_4^+$ and $OCN^-$ (Demyk et al. 1998), and therefore are expected to be formed in an earlier evolutionary stage of dark clouds, when temperatures are as low as 10-20 K and the formation routes through non-energetic atom and radical addition surface reactions dominate.

To date, laboratory experiments on the non-energetic surface formation routes of nitrogen-containing species have mainly focused on the formation of ammonia ($NH_3$), hydroxylamine ($NH_2OH$), and various nitrogen oxides (NO, $NO_2$, $N_2O$) (Hiraoka et al. 1995, Hidaka et al. 2011, Congiu et al. 2012a and 2012b, Fedoseev et al. 2012, Minissale et al. 2014, Ioppolo et al. 2014). The present work extends previous studies on the solid-state formation of ammonia to non-polar (CO-rich) ices, and, at the same time, discusses the link between the surface formation of HNCO. In the accompanying paper (Fedoseev et al. 2014), we investigate the deuterium enrichment of all the ammonia isotopologues as produced through the competition between hydrogenation and deuteration of nitrogen atoms. These results are not further discussed in the present paper.

It is commonly believed that in addition to the depletion from the gas phase, where $NH_3$ is produced through a series of ion-molecular reactions (*e.g.*, Herbst & Klemperer 1973 and Scott et al. 1997), ammonia formation proceeds through the sequential addition of three H atoms to a single nitrogen atom on the surface of ice dust grains:

$$N + H \rightarrow NH \quad (1)$$

$$NH + H \rightarrow NH_2 \quad (2)$$

$$NH_2 + H \rightarrow NH_3 \quad (3).$$

Reactions (1)-(3) were first tested at cryogenic temperatures by Hiraoka et al. (1995), who performed a temperature programmed desorption experiment upon hydrogenation of N atoms trapped in a matrix of solid $N_2$. Recently, Hidaka et al. (2011) confirmed the formation of ammonia in a solid $N_2$ matrix at low temperatures. Their laboratory detection of $NH_3$ was made after annealing the ice to 40 K in order to desorb the $N_2$ matrix. So far, studies of N-atom hydrogenation in more realistic and astronomically relevant $H_2O$- and CO-rich ice analogues have not been reported. Under such conditions, the intermediate free radicals, NH and $NH_2$, can potentially react with other molecules or free radicals to form new and more complex species, such as HNCO:

$$NH + CO \rightarrow HNCO \quad (4)$$
$$NH_2 + CO \rightarrow HNCO + H \quad (5).$$

Reaction (4) is exothermic. The reactivity of CO with NH has been investigated in a combined experimental and quantum chemical study by Himmel et al. (2002) *via* photo-induced dissociation of $HN_3$ in a 12 K Ar matrix. In their work, matrix experiments indicated that $NH(^3\Sigma)$ reacts with CO under laboratory conditions to form $HNCO(^1A')$. An activation barrier of ~4200 K was derived by means of CCSD(T), CASSCF, and MP2 calculations carried-out to evaluate geometries and energies at the transition state for this spin-forbidden reaction. Although the value of this barrier could be considered quite high (*e.g.*, the activation barrier for CO + H ~500 K; Fuchs et al. 2009), experiments and simulations yield consistent data when taking into account experimental and computational inaccuracies. Reaction (4) has often been proposed in spectroscopic studies to explain the formation of HNCO in mixed interstellar ice analogues processed by proton or UV radiation (*e.g.*, Raunier et al. 2003, van Broekhuizen et al. 2005 and references therein). However, to the best of our knowledge, there are no studies available from literature on the investigation of reaction (4) with non-energetic input. Reaction (5) is endothermic (4800 K) and therefore is unlikely to occur under cold dense molecular cloud conditions (Nguyen et al. 1996).

Once formed in the ice, $NH_3$ and HNCO may react through

$$HNCO + NH_3 \rightarrow NH_4^+ \, NCO^- \quad (6)$$
$$HNCO + H_2O \rightarrow H_3O^+ \, NCO^- \quad (7)$$

(see Raunier et al. 2003 and Theule et al. 2011, respectively) to form $OCN^-$ and $NH_4^+$ during a later stage of the molecular cloud evolution. Theule et al 2011 found an activation energy barrier for reaction (7) of 3127 K, which is too high to make this reaction important for the conditions and timescales typical for young stellar objects. In a follow-up study, Mispelaer et al. 2012 determined a barrier of 48 K for reaction (6), indicating the latter pathway as the most promising one to form $OCN^-$ and $NH_4^+$. As stated before, $OCN^-$ has been observed, and $NH_4^+$ may have been identified in the solid state, but the focus here is to simulate dense molecular cloud conditions, well before thermal and energetic processing of ices become important. The goal of the present study is to experimentally verify the formation of $NH_3$ through reactions (1)-(3), as well as the formation of HNCO through reaction (4) in an astrochemically representative **ice** and for astronomically relevant temperatures.

## 2 EXPERIMENTAL PROCEDURE

### 2.1 Experimental setup

All experiments (summarised in Table **1**) are performed in an ultra-high vacuum (UHV) setup ($SURFRESIDE^2$), constructed to investigate solid-state atom addition reactions at cryogenic temperatures. The system has been extensively described in Ioppolo et al. (2013), and therefore only a brief description is given here. $SURFRESIDE^2$ consists of three UHV chambers with a room-temperature base-pressure in the range of $10^{-9}$-$10^{-10}$ mbar. A rotatable gold-coated copper substrate is placed in the centre of the main chamber, where gasses are introduced and deposited with monolayer precision onto the substrate surface through two metal deposition lines. A monolayer (ML) corresponds to about $10^{15}$ molecules $cm^{-2}$. The substrate temperature is varied between 13 and 300 K using a He closed-cycle cryostat with an absolute temperature accuracy better than ~2 K. Both of the two other UHV chambers contain an atom beam line and are connected to the main chamber with angles of 45° and 135° with respect to the substrate (see Figs. **1**, **3**, and **4** in Ioppolo et al. 2013). In one atom line a commercially available thermal cracking source (Dr. Eberl MBE-Komponenten GmbH, see Tschersich 2000) is used to generate H/D atoms. In the other atom line a microwave plasma atom source (Oxford Scientific Ltd, see Anton et al. 2000) can be used to generate H/D/N/O atoms or radicals, such as OH. A custom made nose-shape quartz-pipe is placed in between each atom source and the substrate. These pipes are designed in a way that products formed upon thermal cracking (*e.g.*, H from $H_2$) or plasma dissociation (*e.g.*, N from $N_2$) experience at least four collisions with the pipe walls before reaching the substrate. This is done to quench electronically or ro-vibrationally excited states before impacting on the ice. A considerable fraction of non-dissociated molecules (*e.g.*, $H_2$/$D_2$ and $N_2$) are present in the beam. The method to derive atom flux values is described in Ioppolo et al. (2013). We want to stress that the N-atom flux is an effective flux, estimated by measuring the amount of products of a series of barrierless reactions involving N atoms in the solid phase. The H-atom flux used here is an absolute flux. In this case, the amount of H atoms present in the beam is directly measured by the QMS in the gas phase. The latter measurement neglects that not every H atom will stick to the surface of the substrate and therefore will be unavailable for further reactions, and it also does not consider H-atom recombination on the ice surface. The absolute H-atom flux, therefore, is an upper limit for the effective H-atom flux.

Metal shutters separate the atom beam lines from the main chamber. The atom beam sources as well as the molecular dosing lines in the main chamber can be operated independently. This versatile design allows for the sequential (pre-deposition) or simultaneous (co-deposition) exposure of selected interstellar ice analogues to different atoms (*e.g.*, H/D/O/N). In the present study, co-deposition experiments are largely used. The ice composition is monitored *in situ* by means of reflection absorption infrared spectroscopy (RAIRS) in the range between 4000-700 $cm^{-1}$ and with a spectral resolution of 1 $cm^{-1}$ using a Fourier transform infrared (FTIR) spectrometer. The main chamber gas-phase composition is monitored by a quadrupole mass spectrometer (QMS), which is placed behind the rotatable substrate, and is mainly used during temperature programmed desorption (TPD) experiments. Here, RAIRS is used as the main diagnostic tool, complemented with TPD data to constrain the experimental results. Although QMS provides us with a better sensitivity, preference is given to the RAIRS due to the *in-situ* nature of the method.

## 2.2 Performed experiments

The formation of solid $NH_3$ and HNCO is studied for a selected set of well defined experimental conditions. Firstly, all the used gases (CO, $H_2$/$D_2$, and $N_2$) are prepared in distinct pre-pumped (< $10^{-5}$ mbar) dosing lines. Pure $H_2$/$D_2$ gas (Praxair 5.0/Praxair 2.8) is introduced into the tungsten capillary pipe of the thermal cracking source. Pure $N_2$ gas (Praxair 5.0) is dissociated in the plasma chamber of the microwave plasma source. A simultaneous co-deposition of H/D and N atoms with CO gas (Linde 2.0) is performed on the surface of the bare gold substrate, typically at 13 K. RAIR difference spectra are acquired every 5 minutes with respect to the spectrum of the bare gold substrate. For the crucial experiments, once the co-deposition is completed, a new spectrum is taken and used as background reference. Two additional control experiments are then performed on top of the previously grown ice. The first one is a co-deposition of CO molecules with H/D (*i.e.*, without N) atoms and the second one is a co-deposition of CO molecules with N (*i.e.*, without H/D) atoms. These two experiments are performed under exactly the same experimental conditions used for the very first co-deposition experiment in order to allow for a direct comparison. This procedure guarantees that the production of $NH_3$ and HNCO is the cumulative outcome of a low temperature co-deposition of H/D, N, and CO, ruling-out other possible formation pathways due to contaminations in the atom lines or in the main chamber. Co-deposition experiments of H/D + N + CO are repeated a second time and a TPD experiment is performed right afterward to monitor desorption of the formed species by means of the QMS.

A complementary set of control experiments is used to further verify the HNCO formation under astronomically relevant conditions. In this case, pure $NH_3$ vapour is introduced into the microwave plasma source, and the plasma dissociation products (*i.e.*, NH and $NH_2$ radicals together with $NH_3$, H, N, $H_2$ and $N_2$) are co-deposited with CO molecules. During this co-deposition experiment, RAIR difference spectra are acquired every 5 minutes with respect to a spectrum of the bare gold substrate. After completion of the co-deposition, a TPD experiment is performed and desorbing species are monitored by means of the QMS. The presence of NH and $NH_2$ radicals in the beam is verified by performing a co-deposition of $NH_3$ plasma dissociation products with D atoms and observing the N-D stretching mode in the mid-IR. For clarity, in Table **1** only the relevant experiments performed in this study are listed.

**Table 1.** List of the performed experiments.

| Ref. N | Experiment | Ratio | $T_{sample}$ (K) | $R_{dep}$ (ML min$^{-1}$) | Atom-flux$^{TL}$ ($10^{15}$ cm$^{-2}$ min$^{-1}$) | Atom-flux$^{PL}$ ($10^{15}$ cm$^{-2}$ min$^{-1}$) | *t* (min) | *TPD*$^a$ | *Detection of $NH_3$*$^b$ | *Detection of HNCO*$^b$ |
|---|---|---|---|---|---|---|---|---|---|---|
| Verification of $NH_3$ formation | | | | | | | | | | |
| | | | | CO | H (from $H_2$) | N (from $N_2$) | | | | |
| 1.1 | $N{:}H{:}N_2{:}CO$ | 1:20:100:100 | 13 | 0.5 | 0.1 | 0.005 | 60 | - | Y | N |
| 1.2 | $N{:}H{:}N_2{:}CO$ | 1:20:100:100 | 13 | 0.5 | 0.1 | 0.005 | 180 | - | Y | N |
| 1.3 | $N{:}H{:}N_2{:}CO$ | 1:20:100:500 | 13 | 2.5 | 0.1 | 0.005 | 60 | - | Y | N |
| 1.4 | $N{:}H{:}N_2{:}CO$ | 1:100:100:100 | 13 | 0.5 | 0.5 | 0.005 | 60 | - | Y | N |
| 1.5 | $N{:}H{:}N_2{:}CO$ | 1:100:100:100 | 25 | 0.5 | 0.5 | 0.005 | 60 | - | N | N |
| | | | | CO | D (from $D_2$) | N (from $N_2$) | | | | |
| 1.6 | $N{:}D{:}N_2{:}CO$ | 1:20:100:100 | 13 | 0.5 | 0.1 | 0.005 | 60 | - | Y | N |
| | | | | $H_2O$ | H (from $H_2$) | N (from $N_2$) | | | | |
| 2.1 | $N{:}H{:}N_2{:}H_2O$ | 1:20:100:500 | 15 | 2.5 | 0.1 | 0.005 | 90 | - | - | - |
| 2.2 | $N{:}H{:}N_2{:}H_2O$ | 1:20:100:100 | 13 | 0.5 | 0.1 | 0.005 | 60 | - | - | - |
| Verification of HNCO formation through hydrogenation of N atoms in CO-rich ice analogues | | | | | | | | | | |
| | | | | CO | H (from $H_2$) | N (from $N_2$) | | | | |
| 3.1 | $N{:}H{:}N_2{:}CO$ | 1:20:100:100 | 13 | 0.5 | 0.1 | 0.005 | 90 | $QMS^{2K/5K}$ | Y | N |
| 3.2 | $N{:}H{:}N_2{:}CO$ | 1:6:100:100 | 13 | 0.5 | 0.03 | 0.005 | 90 | $QMS^{2K/5K}$ | Y | Y/N |
| 3.3 | $N{:}H{:}N_2{:}CO$ | 1:6:100:100 | 13 | 0.5 | 0.03 | 0.005 | 90 | RAIRS$^c$ | Y | Y/N |
| 3.4 | $N{:}H{:}N_2{:}CO$ | 1:4:100:100 | 13 | 0.5 | 0.02 | 0.005 | 90 | $QMS^{2K/5K}$ | Y | Y |
| 3.5 | $N{:}H{:}N_2{:}CO$ | 1:2:100:100 | 13 | 0.5 | 0.01 | 0.005 | 90 | $QMS^{2K/5K}$ | Y/N | Y |
| 3.6 | $N{:}H{:}N_2{:}CO$ | 1:2:100:100 | 13 | 0.5 | 0.01 | 0.005 | 90 | $QMS^{0.4K/5K}$ | Y/N | Y |
| 3.7 | $N{:}H{:}N_2{:}CO$ | 1:2:100:100 | 13 | 0.5 | 0.01 | 0.005 | 180 | $QMS^{2K/5K}$ | Y/N | Y |
| 3.8 | $N{:}H{:}N_2{:}CO$ | 1:1.5:100:100 | 13 | 0.5 | 0.0075 | 0.005 | 90 | $QMS^{2K/5K}$ | N | Y |
| 3.9 | $N{:}H{:}N_2{:}CO$ | 1:1:100:100 | 13 | 0.5 | 0.005 | 0.005 | 90 | $QMS^{2K/5K}$ | N | Y |
| 3.10 | $N{:}H{:}N_2{:}CO$ | 1:2:100:100 | 25 | 0.5 | 0.01 | 0.005 | 90 | $QMS^{1K/5K}$ | N | N |
| 3.11 | $N{:}H{:}N_2{:}CO$ | 1:1.5:100:100 | 25 | 0.5 | 0.0075 | 0.005 | 90 | $QMS^{2K/5K}$ | N | N |
| Isotope shift experiments confirming the formation of $HN^{13}CO$, $H^{15}N^{13}CO$ and $D^{15}NCO$ | | | | | | | | | | |
| | | | | $^{13}CO$ | H (from $H_2$) | N (from $N_2$) | | | | |
| 4.1 | $N{:}H{:}N_2{:}^{13}CO$ | 1:2:100:100 | 13 | 0.5 | 0.01 | 0.005 | 90 | $QMS^{2K/5K}$ | Y/N | Y |
| 4.2 | $N{:}H{:}N_2{:}^{13}CO$ | 1:2:100:100 | 13 | 0.5 | 0.01 | 0.005 | 360 | RAIRS$^c$ | Y | Y |
| | | | | $^{13}CO$ | H (from $H_2$) | $^{15}N$ (from $^{15}N_2$) | | | | |
| 4.3 | $^{15}N{:}H{:}^{15}N_2{:}^{13}CO$ | 1:2:100:100 | 13 | 0.5 | 0.01 | 0.005 | 90 | $QMS^{2K/5K}$ | Y/N | Y |
| | | | | $^{13}CO$ | D (from $D_2$) | $^{15}N$ (from $^{15}N_2$) | | | | |
| 4.4 | $^{15}N{:}D{:}^{15}N_2{:}CO$ | 1:2:100:100 | 13 | 0.5 | 0.01 | 0.005 | 90 | $QMS^{2K/5K}$ | Y/N | Y/N |
| Formation of HNCO further constrained *via* interaction of CO with $NH_3$ plasma dissociation products | | | | | | | | | | |
| | | | | CO | | $NH_3^{(dissociated)}$ | | | | |
| 5.1 | $CO{:}NH_3^{(dissociated)}$ | nn | 13 | 0.5 | - | n | 90 | $QMS^{2K/5K}$ | - | Y |
| 5.2 | $CO{:}NH_3$ | nn | 13 | 0.5 | - | - | 90 | $QMS^{2K/5K}$ | - | N |
| 5.3 | $CO{:}NH_3^{(dissociated)}$ | nn | 70 | 0.5 | - | n | 90 | $QMS^{5K}$ | - | N |
| | | | | $^{13}CO$ | | $NH_3^{(dissociated)}$ | | | | |
| 5.4 | $^{13}CO{:}NH_3^{(dissociated)}$ | nn | 13 | 0.5 | - | n | 90 | $QMS^{2K/5K}$ | - | Y |
| Confirmation of the presence of $NH_3$ plasma dissociation products in the beam | | | | | | | | | | |
| | | | | | D (from $D_2$) | $NH_3^{(dissociated)}$ | | | | |
| 6.1 | $D{:}NH_3^{(dissociated)}$ | nn | 13 | - | 0.05 | n | 60 | - | - | - |
| 6.2 | $D{:}NH_3$ | nn | 13 | - | 0.05 | - | 60 | - | - | - |
| 6.3 | $NH_3^{(dissociated)}$ | nn | 13 | - | - | n | 60 | - | - | - |

Experiments are performed in co-deposition under different laboratory conditions; different co-deposition ratios are given; Ref. N is the reference number; $T_{sample}$ is the substrate temperature during co-deposition; $R_{dep}$ is the deposition rate of a selected molecule expressed in ML min$^{-1}$ under the assumption that 1 L (Langmuir) exposure leads to the surface coverage of 1 ML; Atom-flux$^{TL}$ is the thermal cracking source atom flux; Atom-flux$^{PL}$ is the MW plasma source atom flux; absolute uncertainties of H/D- and N- fluxes are 50 and 40%, respectively; *t* is the time of co-deposition; *TPD* is the temperature programmed desorption experiment performed afterward with the TPD rate indicated; *Detection of $NH_3$* is the detection of ammonia either by RAIRS or QMS at the end of co-deposition; *Detection of HNCO* is the detection of isocyanic acid at the end of co-deposition; n – the exact $NH_3$ plasma beam composition is not determined, nn – since the exact $NH_3$ plasma beam composition is unknown the co-deposition ratio is not listed.

[a]Two numbers are given for the TPD rate: the first number is the TPD rate that is used below 50 K to gently remove the bulk of $CO/N_2$ ice, the second number is the TPD rate above 50 K. A higher TPD rate above 50 K is used in order to have a higher peak-to-noise ratio in the QMS. Routinely, 1.5 K/min or 2 K/min are used as TPD rates below 50 K. Since no difference is found in the results between the two rates, 2 K/min is indicated everywhere.

[b]**Y/N means that the detection is uncertain**.

[c]gradual warm-up followed by the acquiring of RAIR spectra is used instead of QMS.

## 3 RESULTS AND DISCUSSION

### 3.1 Formation of $NH_3$

A series of co-deposition experiments (see experiments 1.1-1.6 in Table **1**) is performed to simulate the formation of $NH_3$ under dense cold interstellar cloud conditions, *i.e.*, when gas-phase CO has accreted onto the grains and the UV field is still negligible. A RAIR difference spectrum from a co-deposition of $N{:}H{:}N_2{:}CO$ = 1:20:100:100 at 13 K with a total N-atom fluence of $3\cdot10^{14}$ atoms cm$^{-2}$ (±40%) is shown in the large panel of Fig. **1**. $^{12}CO$ (2140 cm$^{-1}$) and $^{13}CO$ (2092 cm$^{-1}$) are both visible. The inset in Fig. **1** shows this experiment in a smaller spectral range (Fig. **1**a), as well as two more N:H:CO co-deposition experiments at 13 K for different mixing ratios (Figs. **1**b-c), and a control experiment with only $NH_3$ and CO molecules co-deposited at 13 K (Fig. **1**d). The same N-atom effective flux ($8 \cdot 10^{10}$ atoms $s^{-1}$ $cm^{-2}$) within a 40% accuracy and total N-atom total fluence are used in the first three experiments. It should be noted that the amount of $N_2$ in the final mixed ice cannot be disregarded, since the N-atom beam comprises a considerable amount of non-dissociated $N_2$ molecules that, unlike $H_2$, can freeze out at 13 K and form a solid layer of ice (Cuppen & Herbst 2007).

The formation of $NH_3$ is confirmed in Fig. **1** (and inset Fig. **1**a) and 1a by the appearance of two absorption features at $\nu_2$ = 975 $cm^{-1}$ and $\nu_4$ = 1625 $cm^{-1}$ (Abouaf-Marguin et al. 1977, Nelander 1984, Koops et al. 1983). In addition, a third feature at $\nu_3$ = 3430 $cm^{-1}$ is observed in the region of the N-H and O-H stretching modes (not shown in the figure). Furthermore, solid $H_2CO$ ($\nu_2$ = 1728 $cm^{-1}$ and $\nu_3$ = 1499 $cm^{-1}$) shows up as a result of CO hydrogenation. Formation of $H_2CO$ by H-atom addition to CO has been previously studied (*e.g.*, Hiraoka et al. 1994, Zhitnikov & Dmitirev 2002, Watanabe et al 2002, Fuchs at al 2009). Ongoing hydrogenation can form solid $CH_3OH$ that is below its detection limit here. A small feature around 1600 $cm^{-1}$ can be assigned to either $H_2O$ impurity or to the aggregate of $NH_3$. The latter assignment is supported by a negligible admixture of $O_2$ or $H_2O$ in the $N_2$ bottle used in the experiments. Small negative peaks in the range between 1350 and 1750 $cm^{-1}$ are water vapour absorptions along the path of the FTIR beam outside the UHV chamber. Unfortunately, these absorptions are still visible in some of the spectra despite the use of a dry air purged system.

Fig. **1**(b) shows the co-deposition of N:H:$N_2$:CO = 1:20:100:500 at 13 K. In this case, the deposition rate of CO is five times higher than in the experiment plotted in Fig. **1**(a). Fig **1**(a) and **1**(b) show the same $NH_3$ final amount, but the $H_2CO$ peaks are more prominent in the spectrum with higher CO abundance. This is expected, since the effective CO surface coverage for the experiment with $N_2$:CO = 100:500 is about 1.7 times higher than for $N_2$:CO = 100:100 (*i.e.*, 50% CO surface coverage for $N_2$:CO = 100:100 vs. 87% CO surface coverage for $N_2$:CO = 100:500). Clearly, $CH_3OH$ abundances are still below the detection limit.

Fig. **1**(c) shows a co-deposition spectrum with an absolute H-atom flux five times higher (N:H:$N_2$:CO = 1:100:100:100) than the one in Fig. **1**a. This results in a further increase of the formed $H_2CO$, consistent with previous work (*e.g.*, Fuchs et al 2009). In contrast with the $H_2CO$ final yield, ammonia absorption features and, therefore, the corresponding formation yield does not increase: the $\nu_4$ total absorbance shows the same value, while the $\nu_2$ total absorption is even 35-40% less intense than the one in Fig. **1**(a). This apparent inconsistency can be explained by the $\nu_2$ mode (symmetrical deformation) being more sensitive to the ice mixture composition than the $\nu_4$ mode (degenerate deformation), particularly with $H_2CO$ around which can form hydrogen bonds with $NH_3$. The fact that the $\nu_2$ mode is significantly more sensitive to environmental changes than the $\nu_4$ mode was also found by Abouaf-Marguin et al. (1977). The latter work shows that when the hydrogen bonds are formed, the position of the $\nu_2$ band is shifted as much as 70-80 $cm^{-1}$ with respect to the position of the monomeric $NH_3$, while this difference is only 10-30 $cm^{-1}$ for the $\nu_4$ mode. Fig. **1** in Hagen & Tielens (1982) further illustrates this for a 10 K CO matrix. In addition, more $H_2$ is expected to be trapped in the growing matrix in the higher H-atom flux experiment, and this will further affect the environment in which $NH_3$ is isolated.

Finally in Fig. **1**(d), a RAIR spectrum of $NH_3$ co-deposited with CO molecules is shown. The total amount of deposited $NH_3$ is 0.3 ML. This number is about the same as for the N-atom total fluence in each of the three aforementioned experiments. A ratio $NH_3$:CO = 1:500 is chosen to reproduce the ratio used in Fig. **1**(b). The total absorbance of deposited $NH_3$ molecules in Fig. **1**(d) is about 10% and 40% higher for the $\nu_4$ and $\nu_2$ modes, respectively, compared to the abundances of $NH_3$ formed by N-atom hydrogenation in the experiments depicted in Figs. **1**(a) and **1**(b). Although these differences are within the flux uncertainties, the larger difference in the $\nu_2$ mode is likely due to the higher sensitivity of the $\nu_2$ mode toward its environment.

In general, Fig. **1** shows that under our experimental conditions the final $NH_3$ yield is determined by the total amount of available nitrogen atoms at the ice surface, as the integrated area of the $\nu_4$ mode stays near constant in all four plots, while for the $\nu_2$ mode this varies significantly in plot c. The experiments indicate that the hydrogenation of the deposited N atoms is a faster and more efficient process than the hydrogenation of CO ice. Very low activation barriers are therefore expected for reactions (1)-(3). This is consistent with N-atom hydrogenation experiments in a solid $N_2$ matrix by Hiraoka et al. (1994) and Hidaka et al. (2011). The Hidaka et al. (2011) experiments are tested under our experimental conditions, *i.e.* a co-deposition of N:H:$N_2$ = 1:20:100 is performed at 15 K, and $NH_3$ formation is also observed in this experiment.

Finally, we performed a few co-deposition (control) experiments of H- and N-atom beams with $H_2O$ instead of CO (see experiments 2.1 and 2.2 in Table **1**). A strong broadening of the $NH_3$ absorbance features due to hydrogen bonds and a considerable overlap of $H_2O$ and $NH_3$ absorption features do not allow for an unambiguous assignment of $NH_3$ peaks in these experiments, TPD using the QMS does not help to overcome the problem since co-desorbing $H_2O$ gives similar m/z numbers to $NH_3$ complicating the assignments. Therefore these experiments will not be further discussed in this Section.

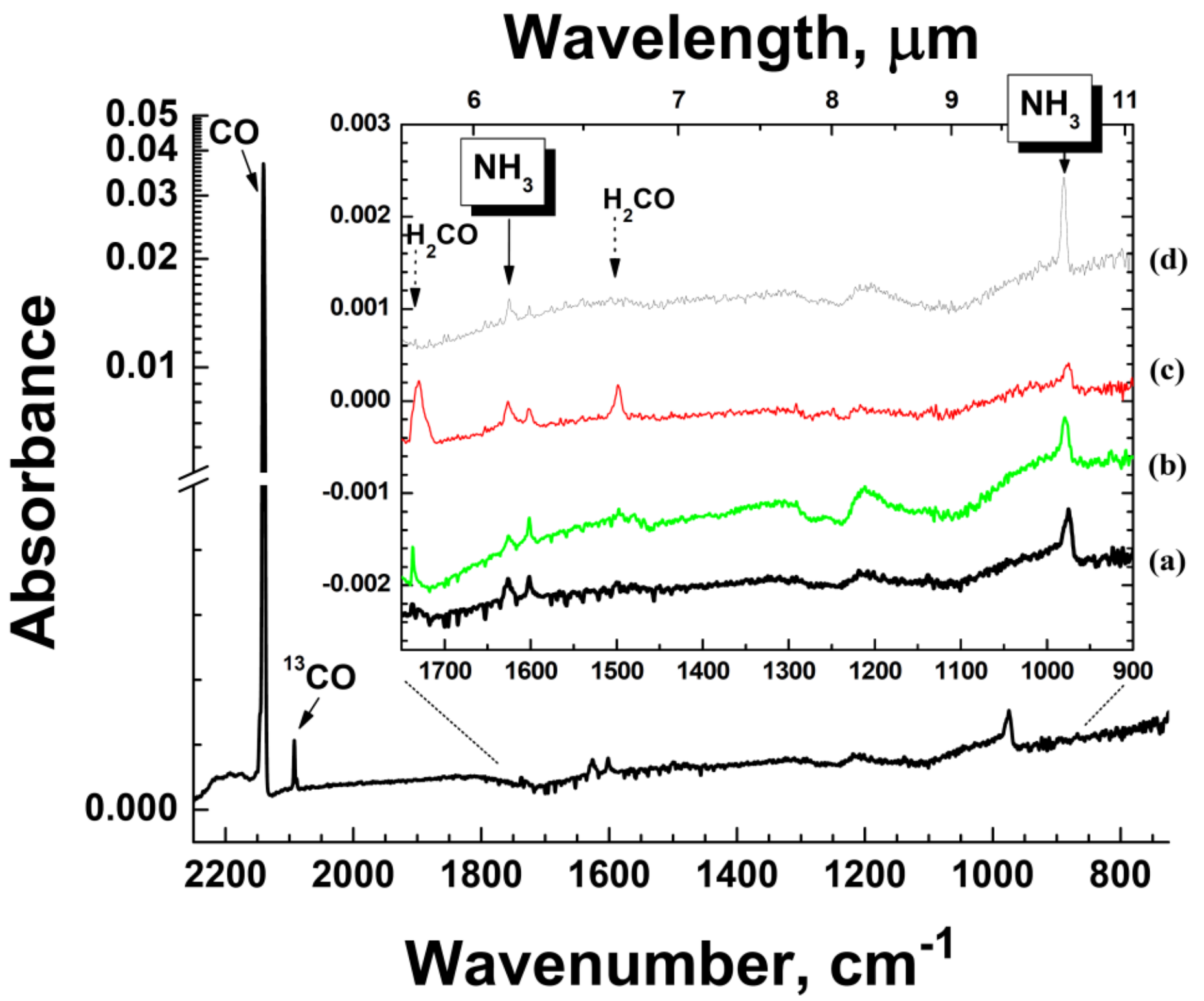

**Figure 1.** A RAIR difference spectrum from a co-deposition of N:H:$N_2$:CO = 1:20:100:100 at 13 K with a total N-atom fluence of $3\cdot10^{14}$ (±40 %) atoms $cm^{-2}$ (experiment 1.1) is shown in the large panel. In the inset four spectra from different co-deposition experiments are shown in a narrower spectral range: (a) is a zoom-in of the aforementioned spectrum; (b) is for a co-deposition of N:H:$N_2$:CO = 1:20:100:500 (experiment 1.3); (c) is for a co-deposition of N:H:$N_2$:CO = 1:100:100:100 (experiment 1.4); and (d) is for the deposition of $NH_3$:CO = 1:500 with a total deposited $NH_3$ amount of 0.3 ML, corresponding to the N-atom total fluence of the experiments (a)-(c). All spectra are for 13 K and plotted with offsets for clarity.

### 3.2 Temperature dependence

A co-deposition experiment using the same deposition rates discussed before ($N:H:N_2:CO$ = 1:100:100:100) is repeated for different substrate temperatures (13 and 25 K) to study the temperature effect on the N-atom hydrogenation in CO-rich ices. The temperatures chosen are below the desorption values of $N_2$, N, and CO molecules (Acharyya et al. 2007). The goal of these experiments is to determine which mechanism is responsible for the formation of $NH_3$ in a CO rich environment, *i.e.*, a Langmuir-Hinshelwood (L-H), Eley-Rideal (E-R) or 'hot-atom' mechanism. Since both E-R and 'hot-atom' mechanisms exhibit very limited sample temperature dependency over the short range of temperatures, one expects to find similar $NH_3$ final yields in both experiments. This kind of dependency is found for example for NO + N co-deposition experiments (Ioppolo et al. 2013). In the case that L-H is responsible for the formation of ammonia, then the resulting $NH_3$ formation rate is a rather complex combination of many individual processes that are temperature dependent (*i.e.*, lifetime of H atoms on the surface, hopping rate of H and N atoms, and H-atom recombination rate). In this case, the $NH_3$ yield should drop significantly at 25 K due to the shorter residence time of H atoms on the ice surface. For instance, the lifetime of H atoms on a water ice surface at 25 K is more than 1000 times shorter than at 13 K (Cuppen & Herbst 2007). A decrease of $H_2CO$ and $CH_3OH$ formation yields with increasing temperature was already observed by Watanabe et al. (2006) and Fuchs et al. (2009) in CO hydrogenation experiments, and this observation was explained assuming a L-H mechanism. Fig. **4** in the accompanying paper (Fedoseev et al. 2014) shows for similar experimental conditions a substantial drop in the amount of ammonia formation between 15 and 17 K further constraining the proposed L-H mechanism.

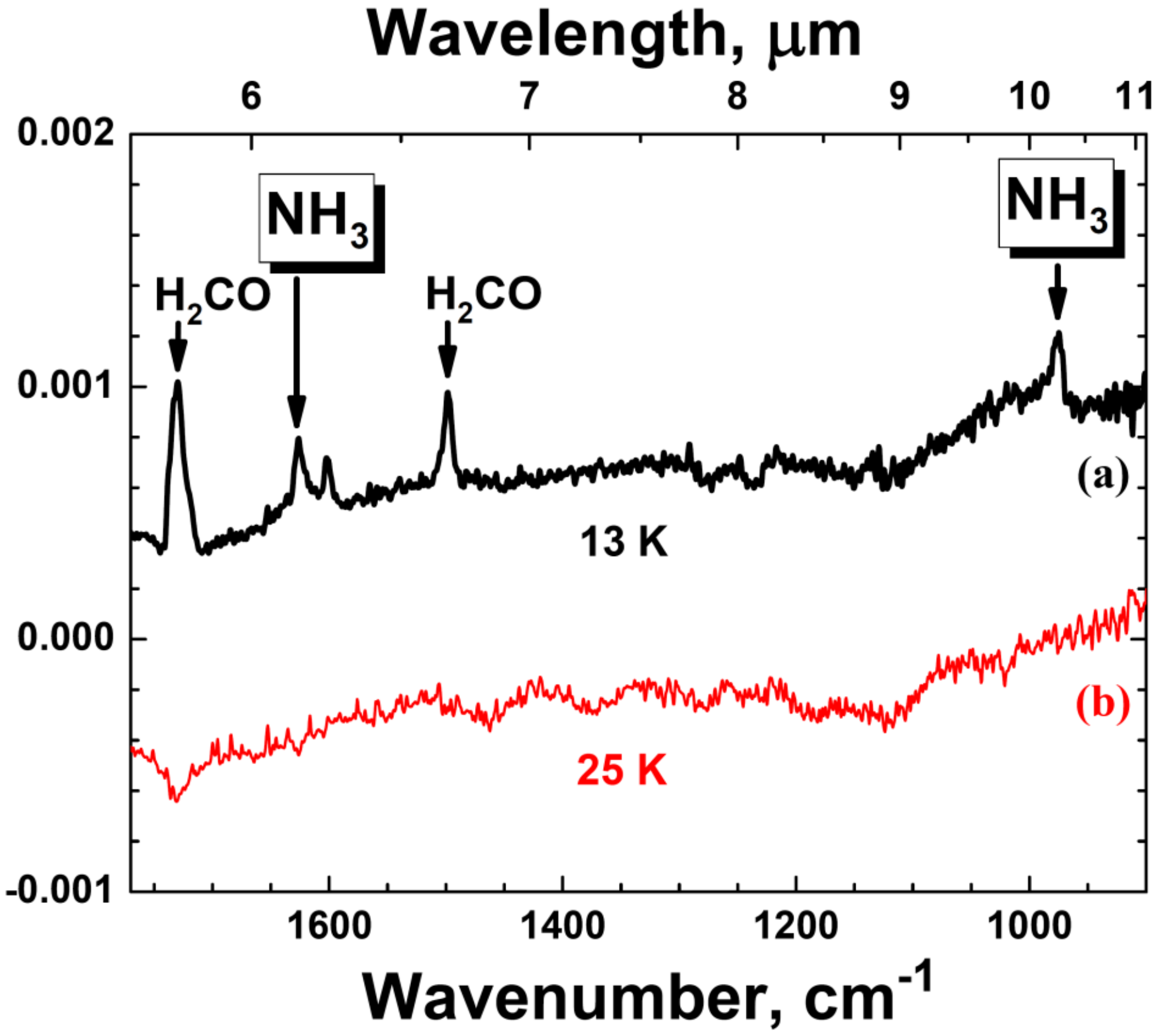

**Figure 2.** Two RAIRs difference spectra of the same co-deposition experiment $N:H:N_2:CO$ = 1:100:100:100 with the same total N-atom fluence of $3 \cdot 10^{14}$ (±40 %) atoms cm$^{-2}$ at 13 K (a) and 25 K (b) (see experiments 1.4 and 1.5). Spectra are plotted with offsets.

The two spectra in Fig. **2** show that neither $H_2CO$ nor $NH_3$ are detected at 25 K by means of RAIRS. This is fully consistent with surface processes following a L-H mechanism, as suggested in previous work that focused on the hydrogenation of N atoms trapped in a $N_2$ matrix (Hidaka et al. 2011).

### 3.3 Formation of HNCO

The surface formation pathway of ammonia through the sequential hydrogenation of N atoms leads to the formation of NH and $NH_2$ intermediates that also can react with other species in the ice to form different molecules. One obvious candidate in a CO surrounding is HNCO formed through surface reactions (4) and (5). In the experiments shown in Figs. **1** and **2**, each NH and $NH_2$ intermediate will face at least one neighbouring CO molecule as with the chosen co-deposition ratios for deposited N atoms there are one hundred CO and one hundred $N_2$ molecules. However, in our RAIR spectra, solid HNCO and its hydrogenation products (*e.g.*, $NH_2CO$ and $NH_2CHO$) cannot be detected. Even sensitive mass spectrometry does not show any clear evidence for the formation of these species (masses 42 – 45 m/z). Thus, reactions (4) and (5) most likely experience an activation barrier and consequently are overtaken by reactions leading to $NH_3$ formation. This is consistent with Himmel et al. (2002), who indeed suggested the presence of an activation barrier even though some reactivity of NH toward CO at 10 K was found.

Here, a new set of experiments is presented to study the surface formation of HNCO through the reaction of CO molecules with NH and $NH_2$. The specific goal of these experiments is to prohibit the (fast) formation of ammonia, and to simultaneously increase the probability for NH and $NH_2$ intermediates to react with CO molecules, overcoming any activation barriers. Such a set of experiments, in fact, is more representative for the actual processes taking place on interstellar grains, where H- and N-atom accretion rates are so low that once NH and $NH_2$ radicals are formed, these experience a relatively long time to react with other ice molecules (~ several days) before another impacting H atom contributes to the formation of ammonia. Thus, to reproduce this scenario, N atoms are co-deposited with CO molecules with the same rates as described in section 3.1, while the H-atom co-deposition rate is substantially decreased (20 times less) to prevent full hydrogenation of N atoms, offering a pathway to the formed NH and $NH_2$ to react with CO (see experiments 3.1-3.11 in Table **1**). TPD experiments combined with QMS data are used to study the expected low HNCO final yield.

Three selected N + H + CO co-deposition experiments are presented in Fig. **3**. After co-deposition of CO molecules with H and N atoms with a given ratio at 13 K, the ice is gently and linearly warmed up to 50 K with a rate of 2 K/min to remove the bulk of the CO ice. A rate of 5 K/min is used during the second part of the TPD (up to 225 K) in order to have a higher peak-to-noise ratio of the selected masses in the mass spectrometer. The correlation between the $NH_3$ and HNCO final yields for different H-atom co-deposition ratios is investigated by integrating the corresponding area of the selected species from their QMS mass signal over time (*i.e.*, m/z = 17 for $NH_3$, and m/z = 42, 43 for HNCO). Fig. **3** shows the decrease of the $NH_3$ formation yield (peak centred at 120 K), and the corresponding gradual increase of the HNCO formation yield (peak centred at 185 K) that follows the decrease of H-atom co-deposition ratio from N:H:$N_2$:CO = 1:20:100:100 to N:H:$N_2$:CO = 1:2:100:100. In the first experiment, only traces of HNCO are detected by the QMS, while the $NH_3$ signal is maximum. In the third experiment, with a 10 times smaller H flux, the HNCO final yield is maximum while only traces of $NH_3$ are present. The intermediate case, corresponding to a co-deposition ratio of N:H:$N_2$:CO = 1:6:100:100 results in the presence of both $NH_3$ and HNCO molecules.

HNCO can be assigned in the TPD experiments to the desorbing species peaked at 185 K by looking at the electron ionization fragmentation pattern (see Bogan & Hand 1971, Fischer et al. 2002). The inset in Fig. **3** compares the ratio between m/z = 43, 42, and 15 (*i.e.*, $HNCO^+$, $NCO^+$, $HN^+$) in our experiment and literature values.

To further constrain this assignment, similar experiments are performed with H atoms co-deposited with $^{14/15}N$ atoms and $^{13}CO$ molecules (see experiments 4.1-4.3 in Table **1**). In both cases ($^{14}N$ and $^{15}N$), a consistent isotopic shift of both peaks at m/z = 42 and 43 is observed, while the ratio between these two peaks stays constant.

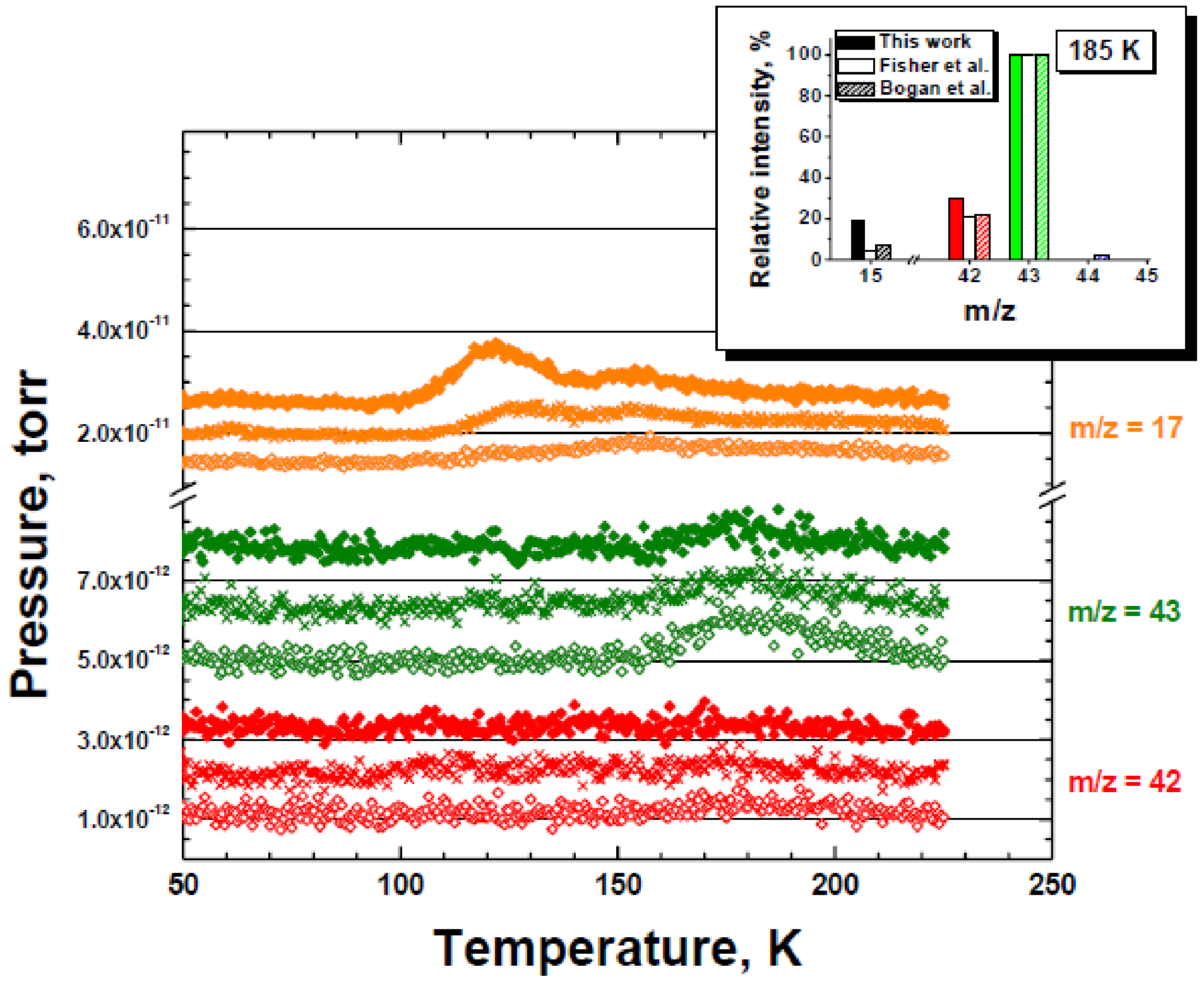

**Figure 3.** The TPD spectra for three different experiments: co-deposition of $N:H:N_2:CO$ = 1:20:100:100 at 13 K (filled diamonds); co-deposition of $N:H:N_2:CO$ = 1:6:100:100 at 13 K (x-crosses); and co-deposition of $N:H:N_2:CO$ = 1:2:100:100 at 13 K (empty diamonds), see experiments 3.1, 3.2, and 3.5 in Table **1**, respectively. The total N-atom fluence in each of the three experiments is $4.5 \cdot 10^{14}$ (± 40 per cent) atom $cm^{-2}$. Peaks at m/z = 17 are due to $NH_3$ (centered at 120 K) and background $H_2O$ (centered at 155 K)**,** m/z = 42 and 43 (middle and lower panel) are the two most intense signals from HNCO. Plots are shown with offsets. In the top right corner, an inset is shown with the relative intensities for m/z = 15, 42, and 43 (HNCO), as derived in this study and compared to the available literature.

Pure HNCO is known to desorb slightly above 120 K (Theule et al. 2011). However, as shown in Fig. **3**, a much higher desorption temperature is found in our experiments. Upon desorption of the bulk of CO ice, HNCO may form ammonium isocyanate ($NH_4^+OCN^-$) or hydronium isocyanate ($H_3O^+OCN^-$) in presence of $NH_3$ or $H_2O$, respectively (reactions (6) and (7)). This indeed shifts the desorption temperature of HNCO to higher values. Reactions (6) and (7) can take place during the thermal processing of mixed $NH_3$:HNCO and $H_2O$:HNCO ices and have been extensively studied by Raunier et al. (2003) and Theule et al. (2011), respectively. Under our experimental conditions, both $NH_3$ and $H_2O$ are present in the ice sample during the TPD: *i.e.*, $NH_3$ is a product of N-atom hydrogenation, and $H_2O$ originates from background deposition (see the second peak around 155 K for m/z = 17 ($OH^+$) in Fig. **3**). In addition, the low final yield of HNCO (< 1 ML) indicates that this molecule likely occupies the surface spots with the highest binding energy. This further shifts the desorption temperature to higher values. Unfortunately, the presence of background water in the main UHV chamber gives both m/z = 18 ($H_2O^+$) and m/z = 17 ($OH^+$). CO that is present in the main chamber after co-deposition gives m/z = 16 signal ($O^+$). As a consequence it is not possible to make unambiguous assignments for the base counter parting HNCO acid. Finally, it should be noted that both $NH_2CHO$ (formamide) and $(NH_2)_2CO$ (urea), two possible chemical derivatives of HNCO and $NH_4^+NCO^-$, respectively, cannot be observed under our experimental conditions and must be under the detection limit of both QMS and RAIRS.

### 3.3.1 Control Experiments

We performed several control experiments to constrain the formation of HNCO and $OCN^-$ at low temperatures by using RAIRS and QMS techniques during TPD experiments (see experiments 3.3 and 4.2 in Table **1**). RAIR spectra can only be used to identify new species formed in the ice when their final yield is > 0.1 ML. To enhance the RAIR signal for species, like HNCO and $OCN^-$, we performed co-deposition experiments two times longer (experiment 3.3) and four times longer (experiment 4.2) than the corresponding experiments shown in Fig. **3**. Unfortunately, the main infrared absorption band of HNCO at 2260 $cm^{-1}$ (Teles et al. 1989) lies too close to the adsorption features of atmospheric $CO_2$ that is present along the path of our IR beam, outside the UHV chamber. In addition, CO infrared features overlap with the strongest $OCN^-$ band (van Broekhuizen et al. 2005), making $OCN^-$ detectable only after desorption of the bulk of CO ice. Therefore, a co-deposition experiment with $^{13}CO$ (experiment 4.2) is shown in Fig. **4**. In this figure, some infrared spectra acquired at different temperatures during TPD are presented in two selected spectral regions. The left panel covers the spectral range where $HN^{13}CO$ should be observed (Teles et al. 1989), while the right panel shows the range where $O^{13}CN^-$ is expected. Solid $HN^{13}CO$ can only be observed at 13 K, while $O^{13}CN^-$ is clearly present in the spectra taken at 35 and 50 K, well before desorption at 185 K. Thus, the present results support the hypothesis that $HN^{13}CO$ is formed already at 13 K, during co-deposition, and as soon as $^{13}CO$ desorbs, $HN^{13}CO$ reacts with $NH_3$ or $H_2O$ to yield $NH_4^+O^{13}CN^-$ and $H_3O^+O^{13}CN^-$, respectively. Formation of $OCN^-$ is also observed upon desorption of the bulk of the ice in experiment 3.3 using regular $^{12}CO$ isomers.

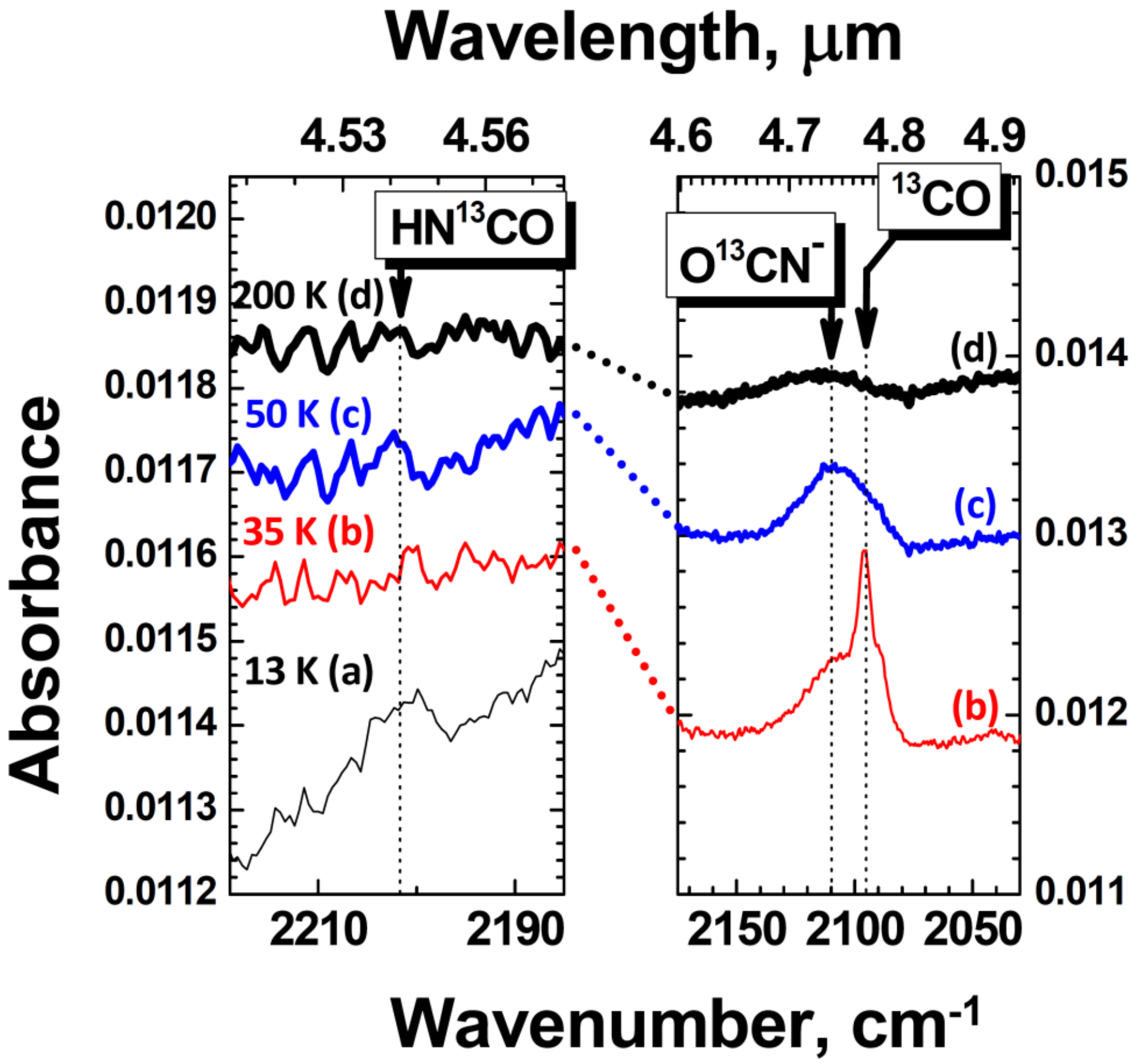

**Figure 4.** Four RAIR difference spectra obtained after co-deposition of N:H:$N_2$:$^{13}CO$ = 1:2:100:100 at 13 K followed by a TPD of the ice (experiment 4.2): a) 13 K b) 35 K c) 50 K and d) 200 K. The left panel shows the strongest absorption feature of $HN^{13}CO$ in the mid-IR; the right panel shows the strongest absorption feature of $O^{13}CN^-$. The total N-atom fluence is $1.8 \cdot 10^{15}$ atoms $cm^{-2}$ (± 40%). In the right panel, the 13 K plot is not shown because of the very high absorption of bulk $^{13}CO$. Some non-desorbed $^{13}CO$ is still present in the plot at 35 K.

Below, additional arguments are discussed that are in favour of the HNCO formation through the interaction of NH/$NH_2$ with CO molecules: (*i*) the co-deposition experiment that yields HNCO at 13 K is repeated for 25 K where it does not result in a QMS detection of HNCO or $NH_3$ (experiments 3.10 and 3.11). This indicates that the involved formation mechanism for both species depends on the H-atom life-time on the ice surface that is known to decrease substantially for increasing co-deposition temperature; (*ii*) a two times longer co-deposition time is applied for identical settings and leads to a two times larger HNCO area on the QMS TPD spectra (experiment 3.7); (*iii*) neither HNNH nor $H_2NNH_2$ are observed (within our detection limits) but both are expected to show up during TPD in the case that non-reacted NH and $NH_2$ become mobile after the bulk of the ice has desorbed.

To further verify the reactivity of NH/$NH_2$ radicals with CO molecules, we performed another set of RAIRS experiments (see experiments 5.1-5.4): a co-deposition of CO molecules with fragmentation products formed by discharging $NH_3$ in the microwave plasma for different experimental conditions. The microwave discharge of ammonia results in the beam containing various plasma dissociation products (*i.e.*, along with the non-dissociated $NH_3$ molecules the beam may contain NH and $NH_2$ radicals, H and N atoms as well as $H_2$ and $N_2$ molecules). The presence of NH and $NH_2$ radicals is confirmed by co-depositing the products of $NH_3$ plasma dissociation with an overabundance of D atoms under the same experimental conditions as in experiments 5.1-5.4. In this case, two broad features are observed in the N-D stretching vibrational mode region (2438 and 2508 $cm^{-1}$, respectively) together with the absorption features due to non-dissociated $NH_3$. This indicates that at least part of the $NH_3$ is decomposed into NH and $NH_2$ radicals and N atoms that then react with D atoms to form the observed N-D bonds.

The QMS TPD spectra obtained after co-deposition of CO molecules with $NH_3$ plasma dissociation products at 13 K results again in a mass peak at m/z = 43 for 185 K that can be assigned to HNCO according to the electron ionisation fragmentation pattern (see filled diamonds in the main panel and the inset in Fig. **5**). Additionally, the empty diamonds in Fig. **5** represent a TPD spectrum after co-deposition of unprocessed $NH_3$ and CO that proves that the formed HNCO is not the result of thermal processing of mixed $NH_3$:CO ice. Cross symbols in Fig. **5** represent data from the experiment where CO molecules are co-deposited with $NH_3$ plasma dissociation products at a substrate temperature of 70 K, well above the CO desorption temperature (peaked at 29 K, Acharyya et al. 2007). This experiment confirms that the formed HNCO does not originate from gas-phase reactions or contaminations in the microwave plasma source and that the presence of CO on the surface is a prerequisite to form HNCO. All the aforementioned experiments are performed by using the same procedure: first the ice is warmed up with a 2 K/min rate from 13 to 50 K to remove the bulk of CO ice; then a rate of 5 K/min is used from 50 K to 225 K. A control experiment is performed with a different isotope (experiment 5.4); when $NH_3$ plasma dissociation products are co-deposited with $^{13}CO$, m/z values of 43 and 44 present a clear feature at 185 K, the ratio 44/43 is 0.27, while m/z = 42 amu is not found.

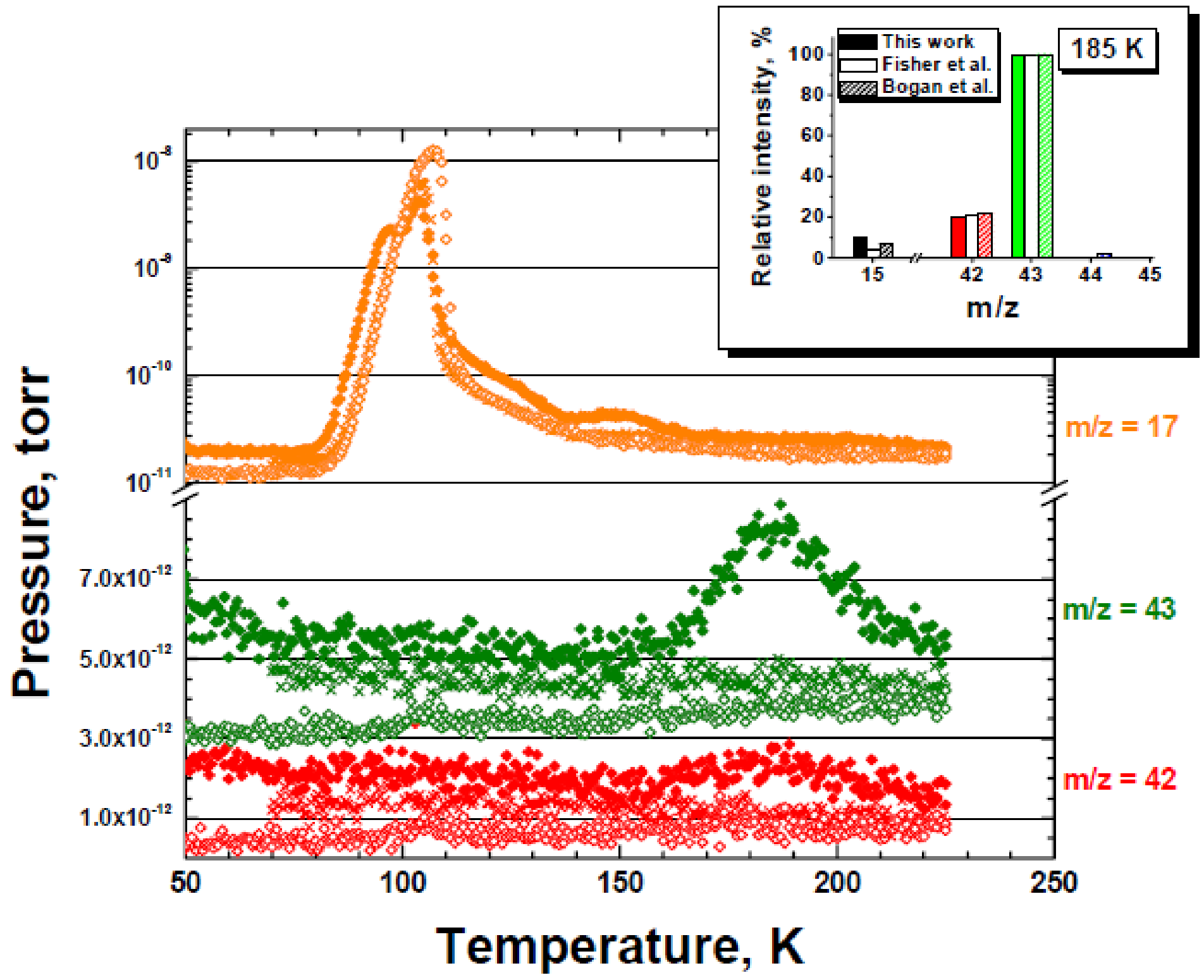

**Figure 5.** The TPD spectra of three distinct experiments: co-deposition of $NH_3$ plasma fragments with CO at 13K (filled diamonds); co-deposition of $NH_3$ plasma fragments with CO at 70 K (x-crosses); and co-deposition of $NH_3$ and CO at 13 K when the microwave discharge is turned off (empty diamonds), see experiments 5.1, 5.3, and 5.2, respectively. Three m/z values are selected: masses 42 and 43 are the two most intense signals from HNCO, while 17 comes from $NH_3$. Plots are shown with offsets for clarity. In the top right corner, the inset compares the relative intensities of the masses assigned in this study to HNCO and the available literature values.

### 3.3.2 HNCO Formation Pathway

Although our experimental data do not allow us to derive values for activation barriers for the reactions (4) or (5), some important conclusions can be drawn**.** As we mentioned before, reaction (5) is endothermic (~4800 K) and therefore unlikely to proceed for non-energetic processing at 13 K. The excess energy of the reactions (1) and (2) does not help to overcome the barrier, because in this case we would observe the formation of HNCO independently from the co-deposition ratio used. Moreover, we would expect a higher HNCO yield for the experiment with a higher H-atom flux over experiments where this flux is insufficient to hydrogenate all N atoms. And this is in contradiction with our experimental observations. Apart from reactions (4) and (5), the following reactions

$$N + CO \rightarrow NCO \quad (8)$$
$$H + NCO \rightarrow HNCO \quad (9),$$

could also lead to the sequential formation of HNCO. However, within our detection limits NCO radicals are not observed after co-deposition of CO molecules with just N atoms. Taking these considerations into account, we conclude that reaction (4) is the main pathway for HNCO formation. Since $H_2CO$ is not detected in the experiments where HNCO is formed, we expect that the activation barrier for the formation of HNCO is not much higher than the one proposed for the H + CO. Fuchs et al. (2009) and Cuppen et al. (2009) used an effective barrier of 435 K to model their observed experimental results on hydrogenation of CO. Such a direct comparison, however, has to be treated with care, as different settings and laboratory conditions have been used and both mobility and life-time of NH and H differ significantly from each other.

As mentioned in section 3.3 $NH_2CHO$ (the product of sequential HNCO hydrogenation) is not observed in any of our experiments. This is not surprising since H-atom addition to HNCO involves an activation barrier. Nguyen et al. 1996 used *ab initio* calculations to study the reaction of H atoms with isocyanic acid and an activation barrier of 1390 K was found for H-atom addition to the nitrogen atom of HNCO. This makes the formation of $NH_2CHO$ unlikely under our experimental conditions, since it would imply a second consequent reaction involving an activation barrier, while the lack of H atoms is used in the experiments resulting in HNCO formation.

## 4 ASTROCHEMICAL IMPLICATIONS

This laboratory work is motivated by several of the main conclusions of the *Spitzer* c2d Legacy ice survey (Öberg et al. 2011). The evolutionary steps of interstellar ice formation can be divided into three main stages: an early phase, driven by fast H-atom addition reactions in cold molecular clouds, before cloud core formation; a later CO freeze-out stage, when chemistry in the ices is driven by accreting CO molecules to a large extent; and the protostellar phase, where thermal and UV processing shape the ice content.

During the first stage, a $H_2O$-rich (polar) ice is formed. In this phase, the relative abundances of $CO_2$ (in $H_2O$), $CH_4$, and $NH_3$ correlate with $H_2O$ ice suggesting their co-formation. This indicates that most of the solid $NH_3$ is formed during an early evolutionary stage. Our laboratory experiments are designed to study the non-energetic surface formation of $NH_3$ through the hydrogenation of N atoms under cold dense cloud conditions. These conditions approximately resemble the first stage of interstellar ice formation. We therefore performed some experiments co-depositing H and N atoms with water, but the spectral confusion due to the overlap of features from $H_2O$ and $NH_3$, as well as the strong broadening of the $NH_3$ absorption bands in a polar ice made an unambiguous assignment of $NH_3$ ice formation far from trivial. Therefore, this work mostly focuses on the investigation of the $NH_3$ formation in non-polar mixtures containing CO ice. This way, newly formed $NH_3$ can be easily identified, because ammonia features are sharper and do not overlap with features of other species in the ice. Although CO-rich ices better represent the second phase of interstellar ices, when CO molecules freeze-out onto the grains, some of our conclusions on the surface formation of ammonia in CO-rich ice can also be extended - within limits - to the formation of $NH_3$ in $H_2O$-rich ice. For instance, we find that the formation of $NH_3$ by hydrogenation of N atoms proceeds barrierless or through a very small activation barrier at 13 K. Moreover, in agreement with previous works, we confirm a Langmuir-Hinshelwood mechanism as the main channel for the formation of solid $NH_3$ (Hiraoka et al. 1995 and Hidaka et al. 2011). Our experimental results further constrain the findings described in Charnley et al. 2001, where the accretion of gas-phase ammonia in their model results in a solid $NH_3/H_2O$ ratio of only ~2%, which is less than the observed values of 5% (Gibb et al. 2004, Bottinelli et al. 2010, Öberg et al. 2011). However, the amount of $NH_3$ ice on the grains could be higher assuming that N atoms also accrete onto grains and undergo hydrogenation. If this surface formation route of ammonia is included in the Charnley et al. (2001) model, the solid $NH_3/H_2O$ ratio becomes ~10%, which is even above the observed value. Moreover, recent models that account for the formation of $NH_3$ in the solid phase through hydrogenation of N atoms indicate a $NH_3/H_2O$ ratio as high as 25% (Garrod & Pauly 2011 and Vasyunin & Herbst 2013).

Our experimental results can explain why recent astrochemical models overestimate the surface production of $NH_3$. We observe an efficient formation of HNCO ice in H+N+CO experiments. Solid HNCO is a product of the interaction between CO molecules and intermediates involved in the surface formation of $NH_3$. In this scenario, $NH_3$ ice is formed efficiently in a polar ice together with water during the first phase of interstellar ices. However, as soon as densities are high enough for CO to freeze-out onto the grains, the formation of $NH_3$ competes with the formation of HNCO in a non-polar ice. Our experiments reveal that the formation of $NH_3$ in CO-rich ices is only efficient when the H-atom deposition rate is high enough to quickly hydrogenate all the N atoms to $NH_3$ which can only occur on rather fast laboratory time-scales. If a slower H-atom deposition rate is used to simulate the slow accretion rate observed in the ISM as much as possible, the formation of $NH_3$ is suppressed in favour of the formation of HNCO. In this case, formed NH or $NH_2$ radicals have significantly more time to overcome the activation barrier of the reaction with the surrounding CO molecules before the next H atom arrives and eventually converts it to $NH_3$. In space, the extremely low accretion rate of H atoms on the surface of the icy grains (unfortunately, not reproducible in the laboratory) gives days to each of the intermediates to overcome activation barriers and to react with the surrounding molecules (*i.e.*, CO ice), before the next hydrogenation event occurs. This potentially explains the observed low ammonia abundances in non-polar ices and, at the same time, shows that solid HNCO is formed in molecular clouds.

Interstellar HNCO was first detected in the Sgr B2 molecular cloud complex by Snyder & Buhl 1972. Since its discovery, HNCO has been detected in different environments, as diverse as dark molecular clouds and hot cores in massive star-forming regions. Li et al. 2013 studied the spatial distribution of HNCO in massive YSOs that is consistent with a 'hot' gas-phase formation route. In cold dark molecular clouds, however, HNCO is expected to be efficiently formed in the solid phase. Quan et al. (2010) were able to reproduce the observed gas-phase abundances of HNCO and its isomers in cold and warm environments using gas-grain simulations, which include both gas-phase and grain-surface routes. Our work shows that HNCO is efficiently formed under dense cold cloud conditions, i.e., in non-polar ices with a reaction pathway that is linked to the formation of ammonia ice and does not require any energetic input, such as UV light or cosmic ray irradiation. Although our work indicates that HNCO should be detectable in CO-rich ices through its anti-symmetric stretch mode at ~2250 $cm^{-1}$, this is not the case; the non-detection of solid HNCO in the ISM can be explained by efficient destruction pathways that include the hydrogenation of HNCO as well as thermal reactions with solid ammonia at temperatures slightly higher than its formation temperature. The latter mechanism is supported by a combined laboratory and modelling study that derives a low (48 K) activation energy barrier for reaction (6) to occur (Mispelaer et al. 2012).

Öberg et al. (2011) reported a close correlation between CO, $CO_2$ (in CO), CO (in $H_2O$), and the XCN band in support of their co-formation during the CO freeze-out stage as well as the identification of $OCN^-$ as a main carrier of the interstellar XCN band. Our TPD experiments shown in Fig. **4** simulate the heating of ice mantles by a newly formed protostar (*i.e.*, the third phase of interstellar ices). This process leads to the formation of $O^{13}CN^-$ detectable from its infrared absorption feature centred at 2202 $cm^{-1}$ (in our experiments we used $^{13}CO$ instead of the regular $^{12}CO$). While $NH_3$ and $HN^{13}CO$ are formed through non-energetic surface reactions at low temperature, the formation of $O^{13}CN^-$ occurs through the interaction of $HN^{13}CO$ with $NH_3$ or $H_2O$ molecules at higher temperatures, when highly volatile species leave the ice. Thus, our results further constrain the assignment of either the entire or a single component (2165 $cm^{-1}$) of the XCN band observed toward numerous YSOs to solid $OCN^-$ and show that $OCN^-$ is successfully formed in the interstellar ices without any UV or cosmic ray processing involved (see Gibb et al. 2004, Bottinelli et al. 2010, Öberg et al. 2011).

The presence of HNCO and $OCN^-$ in interstellar ices during the protostellar phase may be very important for astrobiology. The recent detection of amino acids in comets has boosted efforts to investigate the astrochemical origin of species such as glycine and alanine (Elsila et al. 2009). Since gas-phase routes to form these complex species seem to be inefficient, solid-phase formation pathways offer a strong alternative (Barrientos et al. 2012). During protostar formation, interstellar grains are exposed to thermal, UV, electron, and ion processing that can drastically modify the composition of the icy mantles. Especially in the case of cosmic ray irradiation, the energy released by the ions passing through a material causes the dissociation of hundreds of molecules along their path. These fragments can then recombine forming new and more complex species. Eventually, a complex polymeric refractory residue can be formed. As shown in Figure **6**, HNCO molecules are included as a peptide bond [-(H)N-C(O)-] between any two single amino acid. Moreover, even the simplest peptide, polyglycine, contains nothing but HNCO and $CH_2$ components. Therefore, energetic processing (*e.g.*, UV photolysis and cosmic ray irradiation) of HNCO:$CH_4$ and HNCO:$CH_4$:$CH_3OH$-rich ices can be a possible pathway to form amino acids or peptide fragments. If $OCN^-$ is used in the aforementioned mixtures instead of HNCO, amino acid anions and their fragments can be formed as well. Such experiments will be the focus of a future laboratory study aimed to investigate the formation of the simplest amino acids and peptide fragments, but for now, it is important to conclude that convincing solid state pathways are found that explain the effective formation of the elementary precursor species.

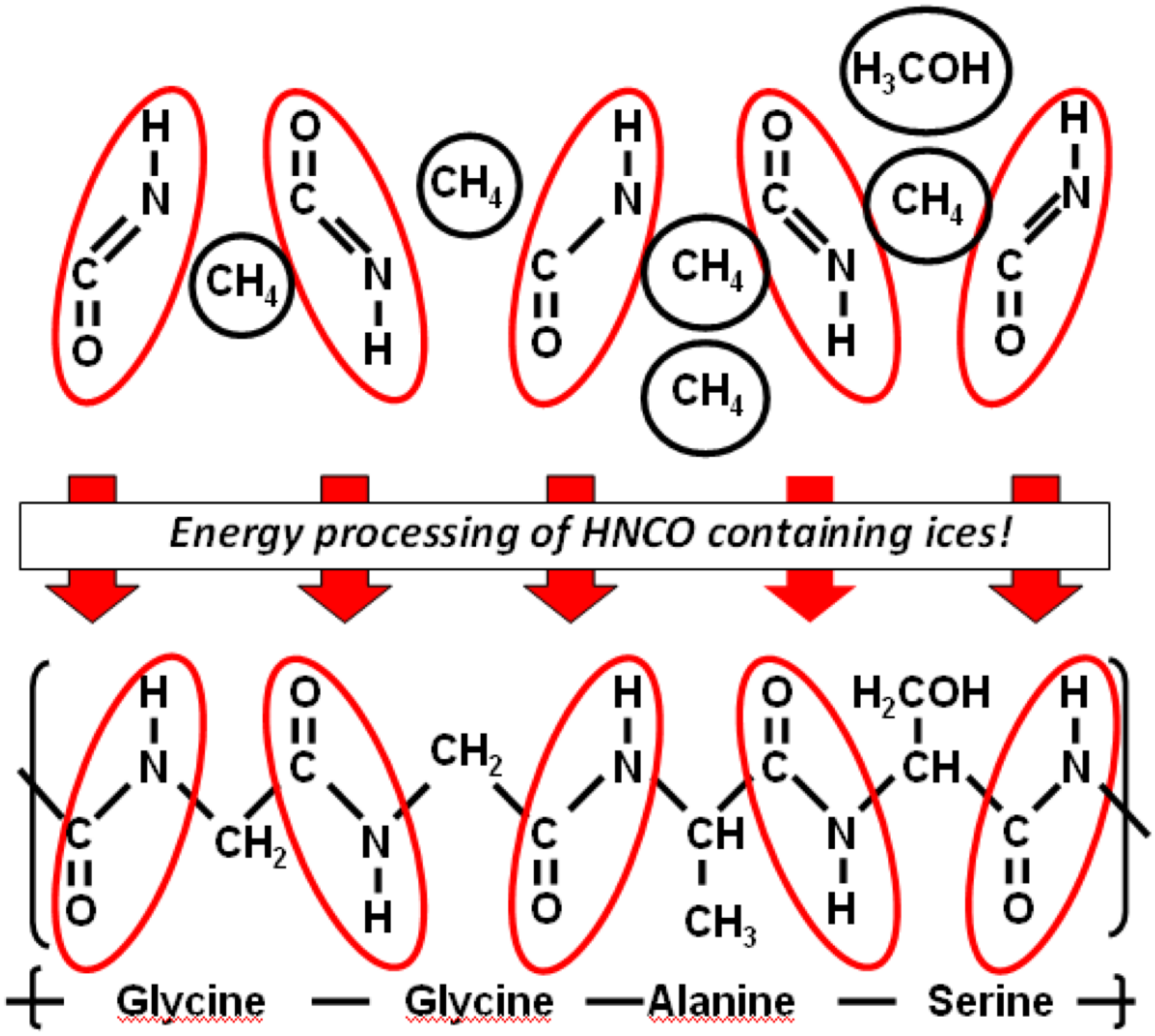

**Figure 6.** A schematic that illustrates the potential importance of HNCO as a simple bearer of peptide bonds for the production of amino acids in interstellar ices.

## ACKNOWLEDGMENTS

We thank Dr. Herma Cuppen for useful discussions. This work is financially supported by the European Community's Seventh Framework Programme (FP7/2007-2013) under grant agreement no. 238258 (LASSIE), the Netherlands Organization for Scientific Research (NWO) through a VICI grant, and NOVA, the Netherlands Research School for Astronomy. Support for S.I. from the Niels Stensen Fellowship and the Marie Curie Fellowship (FP7-PEOPLE-2011-IOF-300957) is gratefully acknowledged.